\documentclass[english]{article}
\usepackage[super,compress]{cite}
\usepackage[T1]{fontenc}
\usepackage[latin9]{inputenc}
\usepackage{babel}
\usepackage{array}
\usepackage{units}
\usepackage{multirow}
\usepackage{amstext}
\usepackage{amssymb}
\usepackage{graphicx}
\usepackage[unicode=true,pdfusetitle,
 bookmarks=true,bookmarksnumbered=false,bookmarksopen=false,
 breaklinks=false,pdfborder={0 0 1},backref=false,colorlinks=false]
 {hyperref}

\makeatletter

\providecommand{\tabularnewline}{\\}

\makeatother

\begin{document}

\title{A Dark Vector Resonance at CLIC}
\author{Constanza Callender\footnote{cony.callender@gmail.com} and Alfonso R. Zerwekh\footnote{alfonso.zerwekh@usm.cl}\\
  Departamento de Física\\
  Universidad Técnica Federico Santa María\\
	and\\
	Centro Científico-Tecnológico de Valparaíso\\
        Valparaíso, Chile
	}

\date{}
\maketitle

\begin{abstract}
One of the main problems in Particle Physics is to understand the
origin and nature of Dark Matter. An exciting possibility is to consider
that the Dark Matter belongs to a new complex but hidden sector. In
this paper, we assume the existence of a strongly interacting dark
sector consisting on a new scalar doublet and new vector resonances,
in concordance with a model recently proposed by our group \cite{VRi2HDM}.
Since in a previous work it was found that it is very challenging
to find the new vector resonances at the LHC, here we study the possibility
of finding them at the a future Compact Linear Collider (CLIC) running
at $\sqrt{s}=3$ TeV. We consider two distinct scenarios: when the non-standard
scalars are heavy, the dark resonance is intense enough to make its
discovery possible at CLIC when the resonance mass is in the range
$[2000,3000]$ GeV. In the second scenario, when the non-standard
scalars are light, the new vector boson is too broad to be recognized
as a resonance and is not detectable except when the mass of the scalars
is close to (but smaller than) a half of the resonance mass and the
scale of the dark sector is high. In all the positive cases, less
than a tenth of the maximum integrated luminosity is needed to reach
the discovery level. Finally, we also comment about the mono-$Z$ production.
\end{abstract}

\section{Introduction}

The accumulated astrophysical evidence in favor of the Cold Dark Matter
hypothesis is, currently, one of the main reasons to study extensions
of the Standard Model (SM). A very popular construction in this area
is to consider a second scalar field transforming as a doublet of
$SU(2)_{L}$, supplemented by an unbroken $Z_{2}$ symmetry. This
is the so called ``Inert Two Higgs Doublet Model'' (i2HDM) \cite{i2HDM-1,i2HDM-2,i2HDM-3}
. Recently, motivated by the hypothesis of a dark matter belonging
to a complex sector with its own set of non-Abelian interactions,
our group studied a modified version of the i2HDM where the standard
sector and the inert scalar doublet transform under different local
$SU(2)$ groups \cite{VRi2HDM}. This setup may arise, for instance,
if the dark matter is a composite state. In this case, the new $SU(2)$
corresponds to a local hidden symmetry that dynamically emerges from
an underlying (but not specified in our effective model) strong interacting
sector. The gauge bosons of this effective $SU(2)$ local group give
origin to the dark vector resonances which are the next massive composite
particles. This construction is a direct analogy of the description
of the pions and the rho meson in usual hadron Physics. 

As we will see in the next section, at low energy, the resulting model
is just the i2HDM with extra massive vector resonances which couple
weakly to the standard sector but strongly to the new scalars. Our
previous study shows that the new resonances introduce important modification
on observables like the predicted relic density of the Dark Matter
candidate \cite{VRi2HDM}. Moreover, we also learned that the eventual
detection of the vector resonances at the LHC is a very challenging
task even in the high luminosity regime. For this reason, it seems
wise to study the detectability of these resonances in future accelerators,
specially in the clean environment of the planned lepton colliders.
However, from the future electron-positron colliders, only CLIC is
planned to run at an energy high enough ($\sqrt{s}=3$ TeV) to explore
the mass range we expect for the vector resonance (overviews of CLIC
and its possibility to search for New Physics can be found in \cite{Strom:2017pjx=00007D,Abramowicz:2016zbo,Abramowicz:2018rjq,Milutinovic-Dumbelovic:2018drv=00007D,vanderKolk:2017urn,Simoniello:2016veu}).
Additionally, the projected luminosity ($2$ ab$^{-1}$) promises
an adequate scenario for discovering subtle effects like the existence
of a dark resonance.  

In this work, we study the possibility of discovering the non-Abelian
vector resonances at CLIC. The paper is organized in the following
way. In section \ref{sec:Model-Reminder}, we briefly describe our
model emphasizing the new vector sector. Then, in section \ref{sec:Results},
we describe our analysis and our results. Finally, in section \ref{sec:Conclusions},
we state our conclusions. 

\section{Model Reminder\label{sec:Model-Reminder}}

Our model is an extension of the Inert Two Higgs Doublet Model and
is based on the gauge group $SU(2)_{1}\times SU(2)_{2}\times U(1)_{Y}$.
Our construction has been completely described in \cite{VRi2HDM}
and here we only remind its most important aspects. 

The standard sector (including a scalar doublet) is assumed to transform
only under $SU(2)_{1}$ and $U(1)_{Y}$ while a second scalar doublet
transforms under $SU(2)_{2}$ (and $U(1)_{Y}$). The breaking down
of the $SU(2)_{1}\times SU(2)_{2}$ to the standard $SU(2)_{L}$ is
described by a non-linear link field $\Sigma$ which transforms as
a bi-doublet under $SU(2)_{1}\times SU(2)_{2}$. The Lagrangian describing
the bosonic sector of the model can be written down as

\begin{eqnarray}
\mathcal{L} & = & -\frac{1}{2}Tr\left[F_{1\mu\nu}F_{1}^{\mu\nu}\right]-\frac{1}{2}Tr\left[F_{2\mu\nu}F_{2}^{\mu\nu}\right]-\frac{1}{4}\left[B_{\mu\nu}B^{\mu\nu}\right]\nonumber \\
 &  & +\frac{u^{2}}{2}Tr\left[\left(D_{\mu}\Sigma\right)^{\dagger}\left(D^{\mu}\Sigma\right)\right]\nonumber \\
 & + & \left(D_{\mu}\phi_{1}\right)^{\dagger}\left(D^{\mu}\phi_{1}\right)+\left(D_{\mu}\phi_{2}\right)^{\dagger}\left(D^{\mu}\phi_{2}\right)-m_{1}^{2}\left(\phi_{1}^{\dagger}\phi_{1}\right)-m_{2}^{2}\left(\phi_{2}^{\dagger}\phi_{2}\right)\nonumber \\
 & + & \lambda_{1}\left(\phi_{1}^{\dagger}\phi_{1}\right)^{2}+\lambda_{2}\left(\phi_{2}^{\dagger}\phi_{2}\right)^{2}+\lambda_{3}\left(\phi_{1}^{\dagger}\phi_{1}\right)\left(\phi_{2}^{\dagger}\phi_{2}\right)\nonumber \\
 & + & \lambda_{4}\left(\phi_{1}^{\dagger}\Sigma\phi_{2}\right)\left(\phi_{2}^{\dagger}\Sigma^{\dagger}\phi_{1}\right)+\frac{\lambda_{5}}{2}\left[\left(\phi_{1}^{\dagger}\Sigma\phi_{2}\right)^{2}+\left(\phi_{2}^{\dagger}\Sigma^{\dagger}\phi_{1}\right)^{2}\right]\label{eq:Lag1}
\end{eqnarray}
where $A_{1\mu}$ and $A_{2\mu}$ are the gauge bosons of $SU(2)_{1}$and
$SU(2)_{2}$ respectively and $B_{\mu\nu}=\partial_{\mu}B_{\nu}-\partial_{\nu}B_{\mu}$
is the field strength tensor of $U(1)_{Y}$. The scalar doublets are
denoted by $\phi_{1}$ and $\phi_{2}$ and their covariant derivatives
are:

\begin{equation}
D_{\mu}\phi_{j}=\partial_{\mu}\phi_{j}-ig_{j}A_{j\mu}\phi_{j}-i\frac{g_{Y}}{2}B_{\mu}\phi_{j}\;\;\mathrm{ with }\;\; j=1,2
\end{equation}
and derivative of the link field is:

\begin{equation}
D_{\mu}\Sigma=\partial_{\mu}\Sigma-ig_{1}A_{1\mu}\Sigma+ig_{2}\Sigma A_{2\mu}.
\end{equation}

The standard fermions, on the other hand, couples to the gauge bosons
of $SU(2)_{1}$ and $U(1)_{Y}$, and to $\phi_{1}$ as in the SM .

After the original symmetry breaking process, which we assume to happen
at a scale $u$ larger than the electroweak scale $v$, the Lagrangian
can be written down in the unitary gauge ($\Sigma=1$) as:

\begin{eqnarray}
\mathcal{L} & = & -\frac{1}{2}Tr\left[F_{1\mu\nu}F_{1}^{\mu\nu}\right]-\frac{1}{2}Tr\left[F_{2\mu\nu}F_{2}^{\mu\nu}\right]-\frac{1}{4}\left[B_{\mu\nu}B^{\mu\nu}\right]\nonumber \\
 & + & \frac{u^{2}}{2}Tr\left[\left(g_{1}A_{1\mu}-g_{2}A_{2\mu}\right)\left(g_{1}A_{1}^{\mu}-g_{2}A_{2}^{\mu}\right)\right]\nonumber \\
 & + & \left(D_{\mu}\phi_{1}\right)^{\dagger}\left(D^{\mu}\phi_{1}\right)+\left(D_{\mu}\phi_{2}\right)^{\dagger}\left(D^{\mu}\phi_{2}\right)+m_{1}^{2}\left(\phi_{1}^{\dagger}\phi_{1}\right)+m_{2}^{2}\left(\phi_{2}^{\dagger}\phi_{2}\right)\nonumber \\
 & - & \lambda_{1}\left(\phi_{1}^{\dagger}\phi_{1}\right)^{2}-\lambda_{2}\left(\phi_{2}^{\dagger}\phi_{2}\right)^{2}-\lambda_{3}\left(\phi_{1}^{\dagger}\phi_{1}\right)\left(\phi_{2}^{\dagger}\phi_{2}\right)\nonumber \\
 & - & \lambda_{4}\left(\phi_{1}^{\dagger}\phi\right)\left(\phi_{2}^{\dagger}\phi_{1}\right)-\frac{\lambda_{5}}{2}\left[\left(\phi_{1}^{\dagger}\phi_{2}\right)^{2}+\left(\phi_{2}^{\dagger}\phi_{1}\right)^{2}\right]\label{eq:Lag2}
\end{eqnarray}

As usual, the electroweak symmetry is spontaneously broken when $\phi_{1}$
acquires a vacuum expectation value (vev) $\left\langle \phi_{1}\right\rangle =(0,v/\sqrt{2})^{T}$.
We assume that $\phi_{2}$ does not get a vev and consequently the
Lagrangian remains invariant under the $Z_{2}$ transformation $\phi_{2}\rightarrow-\phi_{2}$. 

In the limit where the coupling constant associated to $SU(2)_{2}$
($g_{2}$) is much larger than the one associated to $SU(2)_{1}$
($g_{1}$), the physical vector fields can be written in terms of
the gauge eigenstates as follows:

\noindent 
\begin{eqnarray}
A_{\mu} & = & \frac{g_{Y}}{\sqrt{g_{1}^{2}+g_{Y}^{2}}}A_{1\mu}^{3}+\frac{g_{1}g_{Y}}{g_{2}\sqrt{g_{1}^{2}+g_{Y}^{2}}}A_{2\mu}^{3}+\frac{g_{1}}{\sqrt{g_{1}^{2}+g_{Y}^{2}}}B_{\mu}\\
Z_{\mu} & = & -\frac{g_{1}}{\sqrt{g_{1}^{2}+g_{Y}^{2}}}A_{1\mu}^{3}-\frac{g_{1}^{2}}{g_{2}\sqrt{g_{1}^{2}+g_{Y}^{2}}}A_{2\mu}^{3}+\frac{g_{Y}}{\sqrt{g_{1}^{2}+g_{Y}^{2}}}B_{\mu}\\
\rho_{\mu}^{0} & = & -\frac{g_{1}}{g_{2}}A_{1\mu}^{3}+A_{2\mu}^{3}.
\end{eqnarray}

and

\begin{eqnarray}
W_{\mu}^{\pm} & = & A_{1\mu}^{\pm}+\frac{g_{1}}{g_{2}}A_{2\mu}^{\pm}\\
\rho_{\mu}^{\pm} & = & -\frac{g_{1}}{g_{2}}A_{1\mu}^{\pm}+A_{2\mu}^{\pm}
\end{eqnarray}

where $\rho_{\mu}^{0,\pm}$ designates the new vector resonances and
as usual, $A_{n\mu}^{\pm}=\frac{1}{\sqrt{2}}\left(A_{n\mu}^{1}\mp i A_{n\mu}^{2}\right)$.
In the same limit, the masses of the vector states can me expressed as:

\begin{eqnarray}
M_A&=&0 \qquad \mathrm{(exact)}\\
M_Z&\approx& \frac{v\sqrt{g_{1}^{2}+g_{y}^{2}}}{2}\left[1-\frac{1}{2}\frac{g_{1}^{4}}{g_{2}^{2}(g_{1}^{2}+g_{y}^{2})}\right]\\
M_{\rho^0}&\approx& \frac{a v g_2}{2}\left[1+\frac{g_1^2}{2g_2^2} \right]\\
M_{W}&\approx&\frac{v g_1}{2}\left[1-\frac{g_1^2}{2g_2^2} \right]\\
M_{\rho{\pm}}&\approx& \frac{a v g_2}{2}\left[1+\frac{g_1^2}{2g_2^2} \right]
\end{eqnarray}

Notice that the interaction of the dark sector with the standard one
is suppressed by a factor $g_{1}/g_{2}$. Consequently, large values
of $g_{2}$, and vector resonances with masses above $2$ TeV, make
the new sector invisible at current LHC searches \cite{Limit-LHC-1,Limit-LHC-2,Limit-LHC-3}.

In the scalar sector, the spectrum is straightforward since no mass
mixing term arise due to the $Z_{2}$ symmetry. Consequently, near
the minimum of the potential, the scalar doublets can be parameterized
as:

\begin{equation}
\phi_{1}=\frac{1}{\sqrt{2}}\left(\begin{array}{c}
0\\
v+H
\end{array}\right)\qquad\phi_{2}=\frac{1}{\sqrt{2}}\left(\begin{array}{c}
\sqrt{2}h^{+}\\
h_{1}+ih_{2}
\end{array}\right)
\end{equation}
Notice that the $Z_{2}$ symmetry makes the lightest new scalar (which we assume to be
$h_{1}$) stable and a dark matter candidate \cite{VRi2HDM}. For
this reason, we call the new sector (formed by $\rho_{\mu}^{0,\pm}$,
$h_{1}$, $h_{2}$ and $h^{\pm}$) the ``dark sector''.

The model described above has several new parameter such as the masses
of the new vector and scalar states, the scale $u$ and the parameters
of the scalar potential. However, for this work, the only relevant
free parameters of the model are: the masses of the vector resonances
($M_{\rho}$), the masses of the new scalars ($m_{h1}$, $m_{h2}$
and $m_{h\pm}$) and $a\equiv u/v$. In what follows we will take
$a=3,4,5$ and $M_{\rho}\in [2,3]$ TeV since in this way we keep $g_{1}/g_{2}\lesssim0.2$ which
is consistent with our level of approximation. The chosen values of $a$ are the same already
consider in our previous work on the dark matter phenomenology  \cite{VRi2HDM}. They are also representative of  low, moderate and high composite scale  in the dark sector, given the values od $M_{\rho}$. 

The model reproduces the observed relic density provided that $M_{h1}<M_{\rho}$

\section{Results\label{sec:Results}}

The aim of this work is to study the possibility of discovering the
new vector resonance $\rho_{\mu}$ at the future lepton collider CLIC.
The basic idea is to take profit from its high energy mode ($\sqrt{s}=3$
TeV), its expected high integrated luminosity (${\cal L}\approx2$
ab$^{-1}$) and to use the effects of initial state radiation and
radiative return to the resonance in order to scan the relevant range
of possible $\rho$ mass values: $M_{\rho}\in[2,3]$ TeV. The upper
limit of this interval is determined by the maximum center of mass
energy available at CLIC while the lower limit makes us sure that
the resonance has escaped detection at the LHC \cite{Limit-LHC-1,Limit-LHC-2,Limit-LHC-3,VRi2HDM}.
We focus on the process $e{{}^+}e^{-}\rightarrow\mu^{+}\mu^{-}$ which,
at leading order, is described only by the interchange of a photon,
a $Z$-boson and a $\rho^{0}$ in the s-channel.

Our model was implemented in CalcHEP \cite{CalcHEP} using the LanHEP
\cite{LanHEP-1,LanHEP-2} package. We used CalcHEP to generate events
taking into account the initial state radiation with the accelerator
parameters informed by the Particle Data Group \cite{PDG} and listed
in the Table \ref{tabla:ClicParameters}.

\begin{table}[h]
\caption{CLIC parameters}
\begin{center}	
{\begin{tabular}{|c|c|}
\hline 
Parameter & Value\tabularnewline
\hline 
\hline 
Maximum Beam Energy & 1.5 TeV\tabularnewline
\hline 
Bunch length & $4.4\times10^{-3}$ cm\tabularnewline
\hline 
\multirow{2}{*}{Beam Radius} & H: $4.5\times10^{-2}$$\mu$m\tabularnewline
\cline{2-2} 
 & V: $9\times10^{-4}$ $\mu$m\tabularnewline
\hline 
Particles per bunch & $0.37\times10^{10}$\tabularnewline
\hline 
Luminosity  & $6\times10^{34}$ cm$^{-2}$s$^{-1}$\tabularnewline
\hline 
\end{tabular}}
\end{center}
\label{tabla:ClicParameters}
\end{table}

In this simulation, we considered the contributions of the standard
sector as well as the production of the dark resonance. We smeared the momenta of the events
generated with CalcHEP, using a Gaussian distribution, in order to take into account
a finite momentum resolution of the detector. For this purpose, we
use $\nicefrac{\Delta p}{p}=0.05$. Then, we computed the invariant
mass of the $\mu^{+}\mu^{-}$ pair in an interval around the (expected)
mass of the resonance. Finally we fit the spectrum with a Gaussian
resonance and a quadratic background in order to obtain the number
of resonant events and the number of background events. All our simulations
were performed at leading order and we considered only the irreducible
background. An additional limitation of our methodology is that we consider only
a local statistical analysis of the resonance in the sense that the computed statistical
significance of the signal refers only to the local significance and not to the global one.

As was already pointed out in \cite{VRi2HDM}, the possibility of
discovering the new vector bosons in a resonant process depends on
whether they can decay or not into the new scalars. The reason behind
this feature is that we are assuming that the coupling constant $g_{2}$
is large (in order to guarantee small interactions with the standard
sector), making the ``dark sector'' strongly coupled. Consequently
two kinematic regimes open up depending of whether the masses new
scalars are larger or smaller than $M_{\rho}/2$. We call them the
Heavy Scalars and Light Scalars scenarios, respectively. In fact,
the partial decay width of $\rho^{0}$ (which are relevant for the
process we are studying) are given by:

\begin{eqnarray}
\Gamma\left(\rho^{0}\rightarrow\bar{f}f\right) & = & N_{c}\frac{a^{2}M_{W}^{4}}{24\pi v^{2}M_{\rho}^{4}}\left(M_{\rho}^{2}-m_{f}^{2}\right)\sqrt{M_{\rho}^{2}-4m_{f}^{2}}\\
\Gamma\left(\rho^{0}\rightarrow h_{1}h_{2}\right) & = & \frac{\left[M_{\rho}^{2}-\left(m_{h1}+m_{h2}\right)^{2}\right]^{3/2}}{48\pi a^{2}v^{2}M_{\rho}^{3}}\left[M_{\rho}^{2}-\left(m_{h1}-m_{h2}\right)^{2}\right]^{3/2}\\
\Gamma\left(\rho^{0}\rightarrow h^{+}h^{-}\right) & = & \frac{1}{48\pi a^{2}v^{2}}\left[M_{\rho}^{2}-4m_{h\pm}^{2}\right]^{3/2}\\
\Gamma\left(\rho^{0}\rightarrow ZH\right) & \approx & \frac{a^{2}M_{W}^{4}}{48\pi v^{2}M_{\rho}}
\end{eqnarray}

where $f$ represents a standard fermion, $N_{c}$ is the number of
colors and the last equation is given for $M_{W}\ll M_{\rho}$.

As an example, in Figure \ref{Fig:width} we plot the total decay
width of $\rho^{0}$ (with $M_{\rho}=2500$ GeV) for three different
values of $a$ and as a function of $m_{h1}$ when all the non-standard
scalars are degenerated. We clearly see how quickly the total width
grows for $m_{h1}<M_{\rho}/2$, originating the two kinematic regimes. 
\begin{center}
\begin{figure}
\begin{centering}
\includegraphics[scale=0.4]{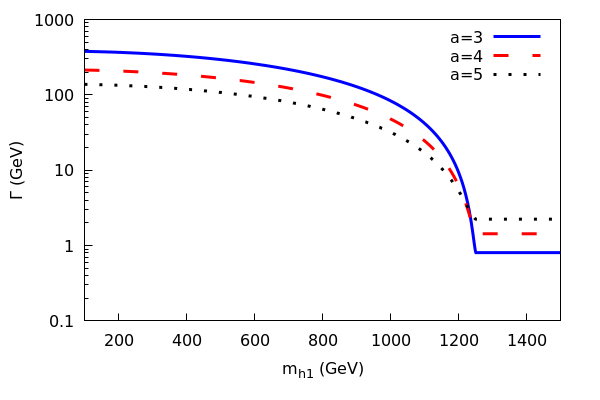}
\par\end{centering}
\caption{Width of the vector resonance ($\rho_{\mu}$) as a function of the
mass of the non-standard scalar in the case they are degenerated.
Here we take $M_{\rho}=2.5$ TeV}

\label{Fig:width}
\end{figure}
\par\end{center}

\subsection{Case 1: Heavy Scalars Scenario}

\begin{figure}
\begin{minipage}[t]{0.5\columnwidth}%
\includegraphics[scale=0.25]{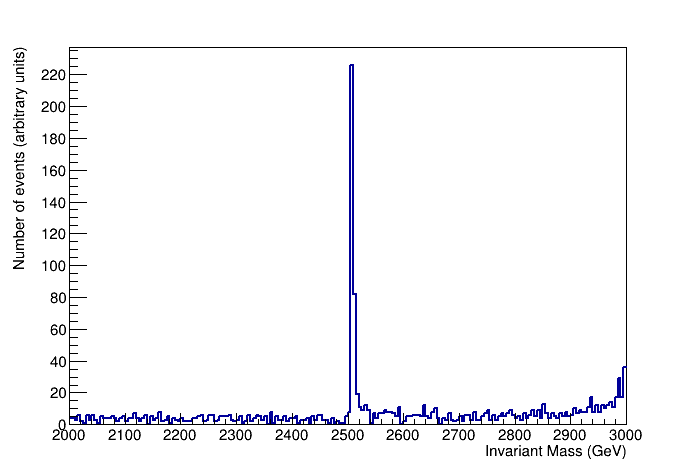}
\begin{center}
a)
\par\end{center}%
\end{minipage}%
\begin{minipage}[t]{0.5\columnwidth}%
\includegraphics[scale=0.25]{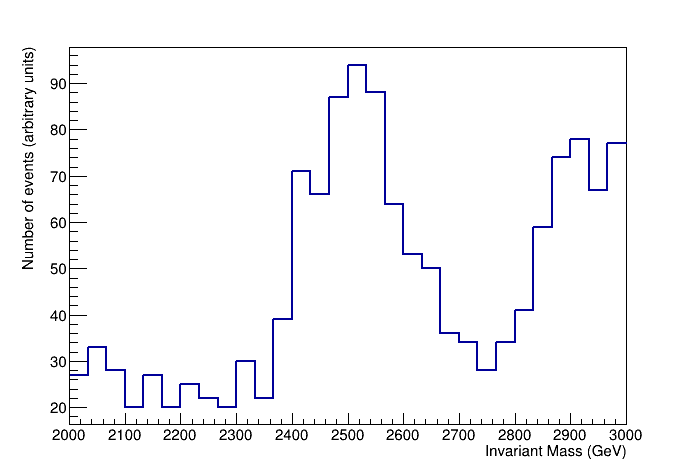}
\begin{center}
b)
\par\end{center}%
\end{minipage}

\caption{Example of resonace in the di-muon invariant mass spectrum with (right)
and without (left) smearing}

\label{Fig:HeavySpectrum}
\end{figure}

\begin{figure}
\begin{minipage}[t]{0.5\columnwidth}%
\includegraphics[scale=0.4]{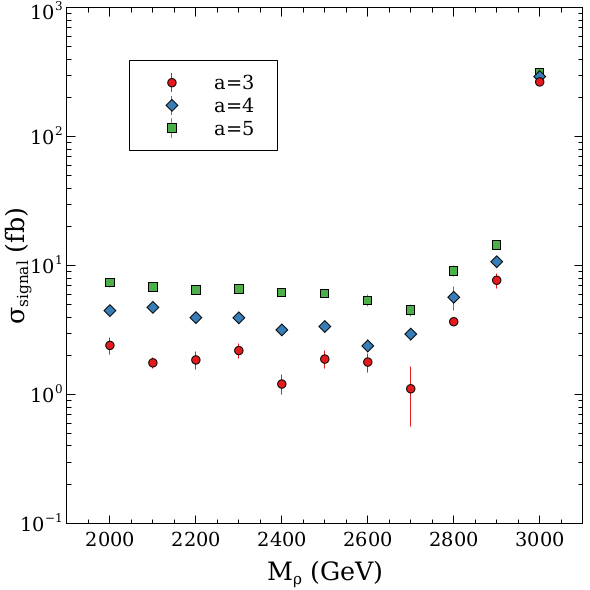}
\begin{center}
a)
\par\end{center}%
\end{minipage}%
\begin{minipage}[t]{0.5\columnwidth}%
\includegraphics[scale=0.4]{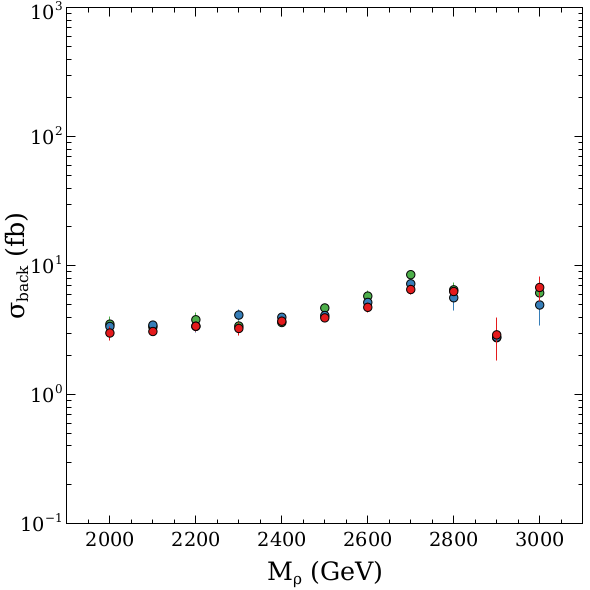}
\begin{center}
b)
\par\end{center}%
\end{minipage}

\caption{Signal (left) and background (right). The error bars reflect the variability
introduced by our Monte Carlo and smearing procedures.}

\label{Fig:SignalAndBackHeavy}
\end{figure}

In the case of heavy scalars, the $\rho^{0}$ appears as a narrow
resonance (with a width of just a few GeV) in the di-muon spectrum
(see Figure \ref{Fig:HeavySpectrum}a). After the smearing procedure,
a broader peak is obtained (Figure \ref{Fig:HeavySpectrum}b). We
fit this peak in the interval $\left[M_{\rho}-\nicefrac{M_{\rho}}{10},M_{\rho}+\nicefrac{M_{\rho}}{10}\right]$
using a Gaussian function (signal) plus a quadratic polynomial (background).
Then we extract a number of events for the signal and a number of
events for the background integrating the Gaussian function and the
quadratic polynomial respectively in the interval defined above .
The corresponding cross sections are shown in Figure \ref{Fig:SignalAndBackHeavy}.
The error bars for the signal and the background reflect the uncertainties
introduced by our Monte Carlo, the smearing of the momenta and our
method of extracting the signal through a fitting procedure. On the
other hand, although the signal and background differential cross
sections decrease with the di-muon invariant mass ($M_{\mu\mu}$),
the energy distribution of the electrons increase with $\sqrt{s}$
reaching its maximum at the maximum nominal center of mass energy
$\sqrt{s_{\mathrm{max}}}=3$ TeV. The reason for this behavior is 
the fact that the events with low energy electrons are due to the 
initial state radiation. Additionally, the interval where the resonance is fitted and integrated
increase with $M_{\rho}$. All these effects produce the $M_{\rho}$
dependence of the signal and background cross sections observed in
Figure \ref{Fig:SignalAndBackHeavy}. Notice that in Figure \ref{Fig:SignalAndBackHeavy}b)
the background obtained for different sets of data (different values
of $a$ ) almost coincide. This feature shows the self-consistency
of our fitting procedure.

\begin{figure}
\begin{centering}
\includegraphics[scale=0.4]{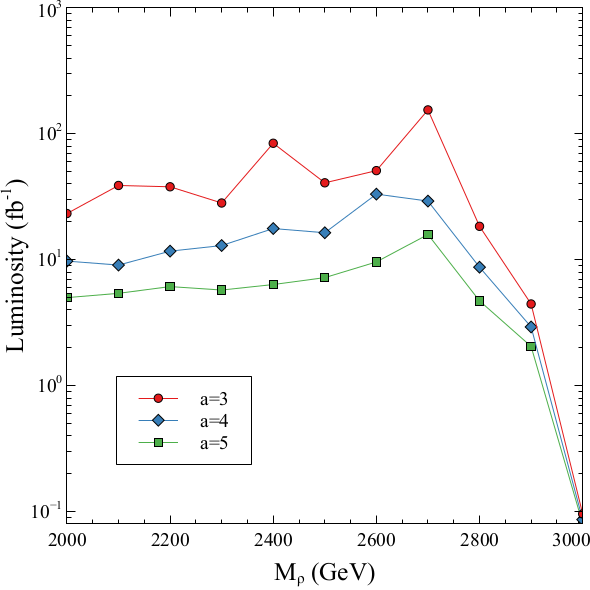}
\par\end{centering}
\caption{Luminosity needed to obtain a (local) significance=5.}

\label{Fig:LuminosityHeavy}
\end{figure}

We see that the signal (Fig. \ref{Fig:SignalAndBackHeavy}a) and the
background (Fig. \ref{Fig:SignalAndBackHeavy}b) cross sections are
of the same order of magnitude and the expected excess of events (for
the projected luminosity) should be clearly observable. Indeed, using
the usual definition for the statistical significance 

\begin{equation}
S=\frac{{\cal L}\sigma_{\mathrm{signal}}}{\sqrt{\mathcal{L}\sigma_{\mathrm{signal}}+\mathcal{L}\sigma_{\mathrm{background}}}},
\end{equation}
 we estimate the luminosity needed to get the discovery criteria $S=5$.
The result is shown in Figure \ref{Fig:LuminosityHeavy}. Notice that the needed luminosity is always much smaller than the maximum luminosity expected at CLIC: $2000$ fb$^{-1}$.


\subsection{Case 2: Light Scalars Scenario}

\begin{figure}
	\begin{centering}
		\includegraphics[scale=0.4]{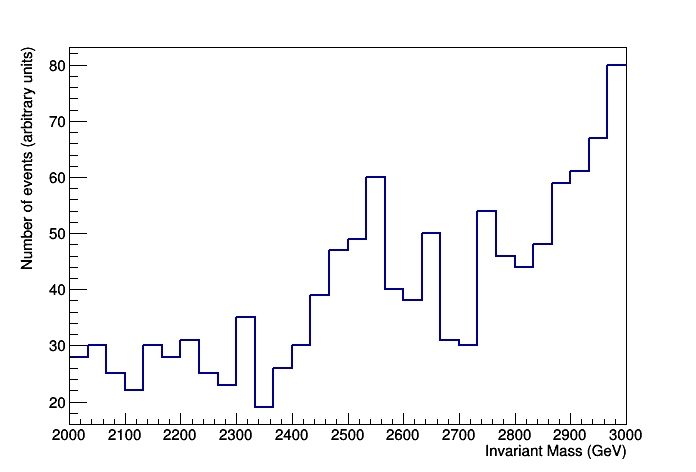}
		\par\end{centering}
	\caption{Example of resonance obtained in the case of light scalars. Here we
		take $M_{\rho}=2500$ GeV, $m_{h1}=m_{h2}=m_{h\pm}=1150$ GeV and
		$a=5$ }
	
	\label{Fig:Resonance_LS}
	
\end{figure}

As we explained above, when the non-standard scalars are light the
$\rho{{}^0}$ becomes broad and, in general it is difficult to identify
it as a proper resonance. Indeed, after the smearing process, we could fit resonances only in the
case of $a=5$ and for this scenario we restrict only to this value of $a$. An example of such a resonance is shown in Figure
\ref{Fig:Resonance_LS}. 

However, this region of the parameter space is worth to be explored. In doing so we found three important alternatives: 1) when only one neutral scalar (the DM candidate) is the only light scalar, 2) when both neutral scalars are light and the charged ones remain heavy and 3) when all the non-standard scalars are light. In the following paragraphs we will comment all these alternatives. For the convenience of the discussion we will define a new parameter ($\delta m$) as

\begin{equation}
\delta m \equiv  M_{\rho}/2-m_{LS}
\end{equation}

where $m_{LS}$ is the mass of the light scalar.

Similarly to the previous case, we fit the spectrum and we identify
the resonant and background events. The only difference is that, this
time, we fit the resonance (and the background) in the interval $\left[M_{\rho}-\nicefrac{M_{\rho}}{5},M_{\rho}+\nicefrac{M_{\rho}}{5}\right]$.

\subsubsection{Only one light scalar}

When only one scalar is light, the neutral vector resonance ($\rho_{0}$) still cannot decay into non-standard scalars since in our model there is no $\rho h_1 h_1$ nor $\rho h_2 h_2$ interaction terms but only $\rho h_1 h_2$. So this case is equivalent to case analyzed in the previous subsection.

\subsubsection{Two neutral light scalars}
\begin{figure}
	\begin{minipage}[t]{0.5\columnwidth}%
		\includegraphics[scale=0.4]{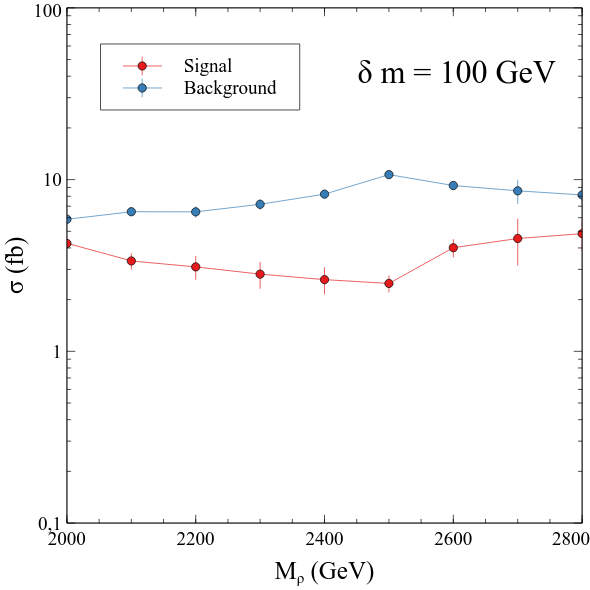}
		\begin{center}
			a)
			\par\end{center}%
	\end{minipage}%
	\begin{minipage}[t]{0.5\columnwidth}%
		\includegraphics[scale=0.4]{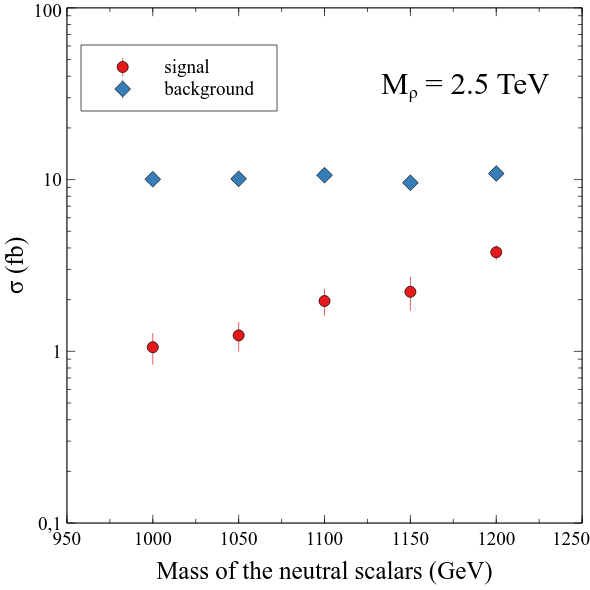}
		\begin{center}
			b)
			\par\end{center}%
	\end{minipage}
	
	\caption{Signal and background cross section for a) $\delta m = 100$ GeV and  $M_{\rho} \in [2,3]$ TeV, and b) $M_{\rho}=2.5$ TeV and masse of the neutral scalars in the range $[950, 1250]$ GeV. In both cases we considered the non-standard neutral scalars degenerated while the charged ones remain heavy. In both plots we use $a=5$.}
	
	\label{Fig:Ndeg}
\end{figure}

When the two neutral scalars are light (but the charged ones remain heavy) the width of the vector resonances receives a moderate increment. For a given value of $M_{\rho}$, we were able to identified resonances provided that $\delta m\lesssim 250$ GeV. In Figure \ref{Fig:Ndeg}, we show the cross sections for signal and background for two situations: a) as a function of $M_{\rho}$ while keeping $\delta m = 100$ GeV and b) as a function of  the mass of the (degenerated) scalars  while keeping $M_{\rho}=2500$ GeV. Although the signal is systematically smaller than the background, the signal over background ratio takes acceptable values ($0.1 \lesssim S/B \lesssim 0.5$). 

\subsubsection{All the scalars are light}

In the following paragraphs, we explore the case where all the scalars
are degenerated
 This degeneration is compatible with the dark matter phenomenology. Indeed our previous study
shows that the model reproduces better the experimental information when the mass difference between $h_1$ and the other scalars is of the order of a few GeV's. Of course, such a small mass difference 
is not important for our collider simulations.

\begin{figure}
\begin{minipage}[t]{0.5\columnwidth}%
	\begin{centering}
		\includegraphics[scale=0.4]{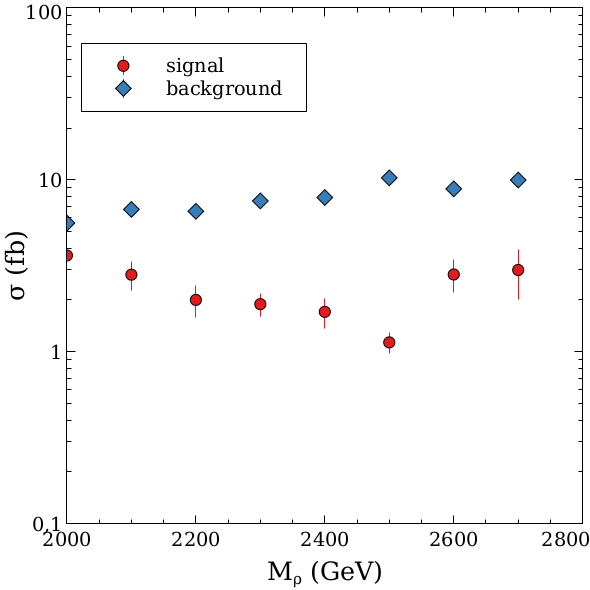}
		\par\end{centering}
	\begin{center}
		a)
		\par\end{center}%
\end{minipage}
\begin{minipage}[t]{0.5\columnwidth}%
	\includegraphics[scale=0.4]{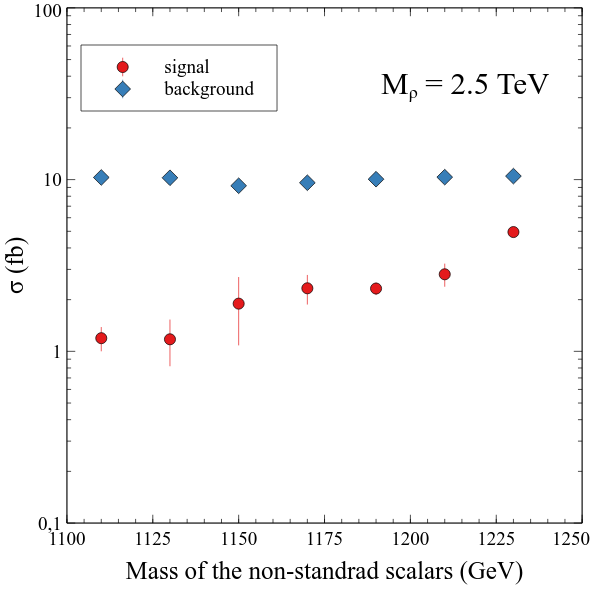}
	\begin{center}
		b)
		\par\end{center}%
\end{minipage}%

\caption{Signal and background cross section for a) $\delta m = 100$ GeV and  $M_{\rho} \in [2,3]$ TeV, and b) $M_{\rho}=2.5$ TeV and masse of the scalars in the range $[1110, 1230]$ GeV. In both cases we considered all the non-standard scalars degenerated. In both plots we use $a=5$.}

\label{Fig:XSctions_LS}

\end{figure}

\begin{figure}
\begin{centering}
\includegraphics[scale=0.4]{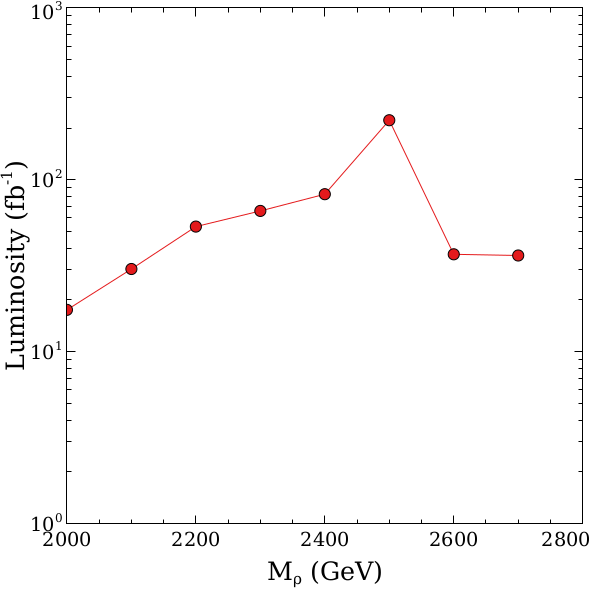}
\par\end{centering}
\caption{Luminosity needed to obtain a significance=5 in the case of light
	scalars and $a=5$}
\label{Fig:CosaTonta}
\end{figure}

In Figure \ref{Fig:XSctions_LS} we show the computed cross sections for the signal and the background for: a) $\delta m = 100$ GeV and  $M_{\rho} \in [2,3]$ TeV, and b) $M_{\rho}=2.5$ TeV and masse of the scalars in the range $[1110, 1230]$ GeV. The signal is again systematically smaller than the background, but still comparable. At the maximum integrated luminosity a few thousands of signal events are expected. As in the previous case, we compute
the integrated luminosity needed for reaching the discovery level.
The results are shown in Figure \ref{Fig:CosaTonta}.

\begin{figure}
	\begin{center}
		\includegraphics[scale=0.4]{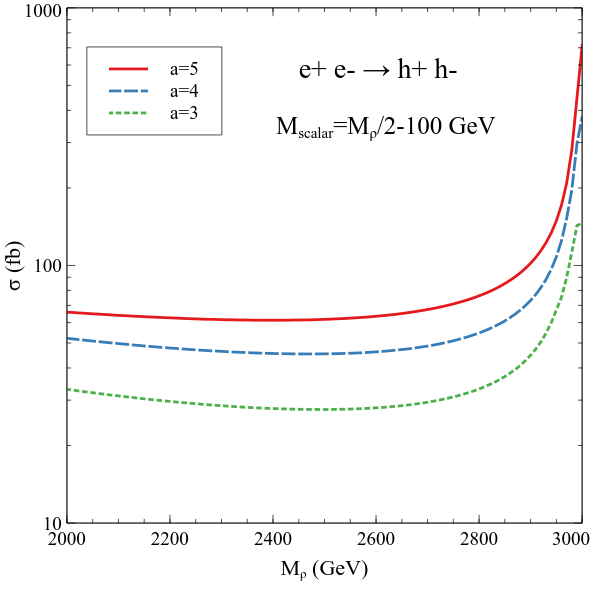}
	\end{center}
\caption{Production cross section for the pair $h^+ h^-$ (in fb) for different values of $M_{\rho}$. Here we have taken the benchmark point $m_{h1}=m_{h2}=m_{h\pm}=M_{\rho}/2 - 100$ GeV}
\label{Fig:eehh}
\end{figure}

\begin{figure}
	\begin{center}
		\includegraphics[scale=0.5]{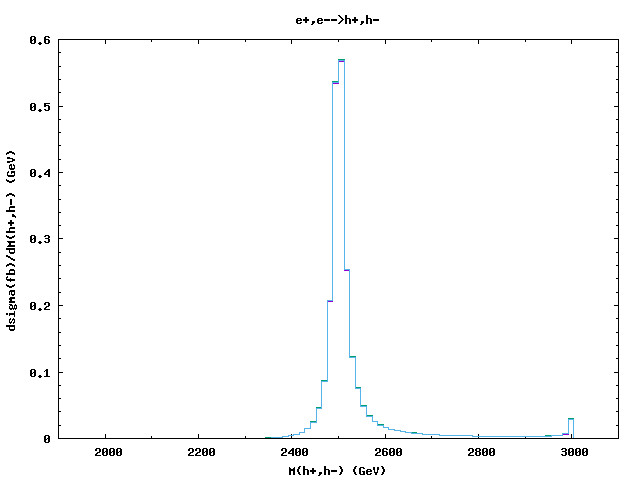}
	\end{center}
	\caption{Invariant Mass distribution of the $h^+ h^-$ pair for $a=3$, $M_{\rho}= 2500$ GeV and  $m_{h1}=m_{h2}=m_{h\pm}=1150$ GeV. We observe the peak corresponding to the resonant production of a $\rho^0$. No smearing has been applied in this case. }
	\label{Fig:hhinvmass}
\end{figure}

In summary, there is some discovery potential (although reduced) in the channel $e^+ e^- \rightarrow \mu^+ \mu^-$ when all the scalars are light, at least when the scale of the vector resonance is high ({\it i.e.} when $a=5$). However, for light scalars a more promising possibility arises through the $e^+ e^- \rightarrow h^+ h^-$ channel. In Figure \ref{Fig:eehh} we plot the production cross section for the pair $h^+ h^-$ for different values of $M_{\rho}$ and assuming that the mass of the scalars is $m_{h1}=m_{h2}=m_{h\pm}=M_{\rho}/2 - 100$ GeV. The all the values of $a$ considered in this paper, we found that the cross sections is significantly  higher in this channel than in the di-muon one. Even the new vector resonance may be observable in the invariant mass distribution of the $h^+ h^-$ as seen in Figure \ref{Fig:hhinvmass}. In this way, the pair production of the charged scalars appears an effective mean of discovering the dark sector at CLIC.

\subsection{Mono-$Z$ Production}

Although in this work our main focus is the study of the di-muon channel
as a mean for discovering the new dark resonance, in the following
paragraphs we will briefly consider the mono-$Z$ production. This channel
is traditionally important for the search of dark matter at colliders.
In our case the main mono-$Z$ production processes are $e{{}^+}e{{}^-}\rightarrow Zh_{1}h_{1}$
and $e\text{\textsuperscript{+}}e\text{\textsuperscript{-}}\rightarrow Zh_{1}h_{2}$,
being the latter the dominant one. For this analysis, we follow an
intermediate scenario where the neutral scalars are supposed light
(in the sense explained above, that is, $m_{h1}=m_{h2}=M_{\rho}/2-100$
GeV ), while the charged scalars are supposed to be heavy. Interestingly,
these two processes are weakly dependent on the $a$ parameter ( in
fact, the resonant production $e{{}^+}e{{}^-}\rightarrow\rho\rightarrow Zh_{i}h_{j}$
is exactly independent of $a$ ).

\begin{figure}
\begin{minipage}[t]{0.5\columnwidth}%
\includegraphics[scale=0.11]{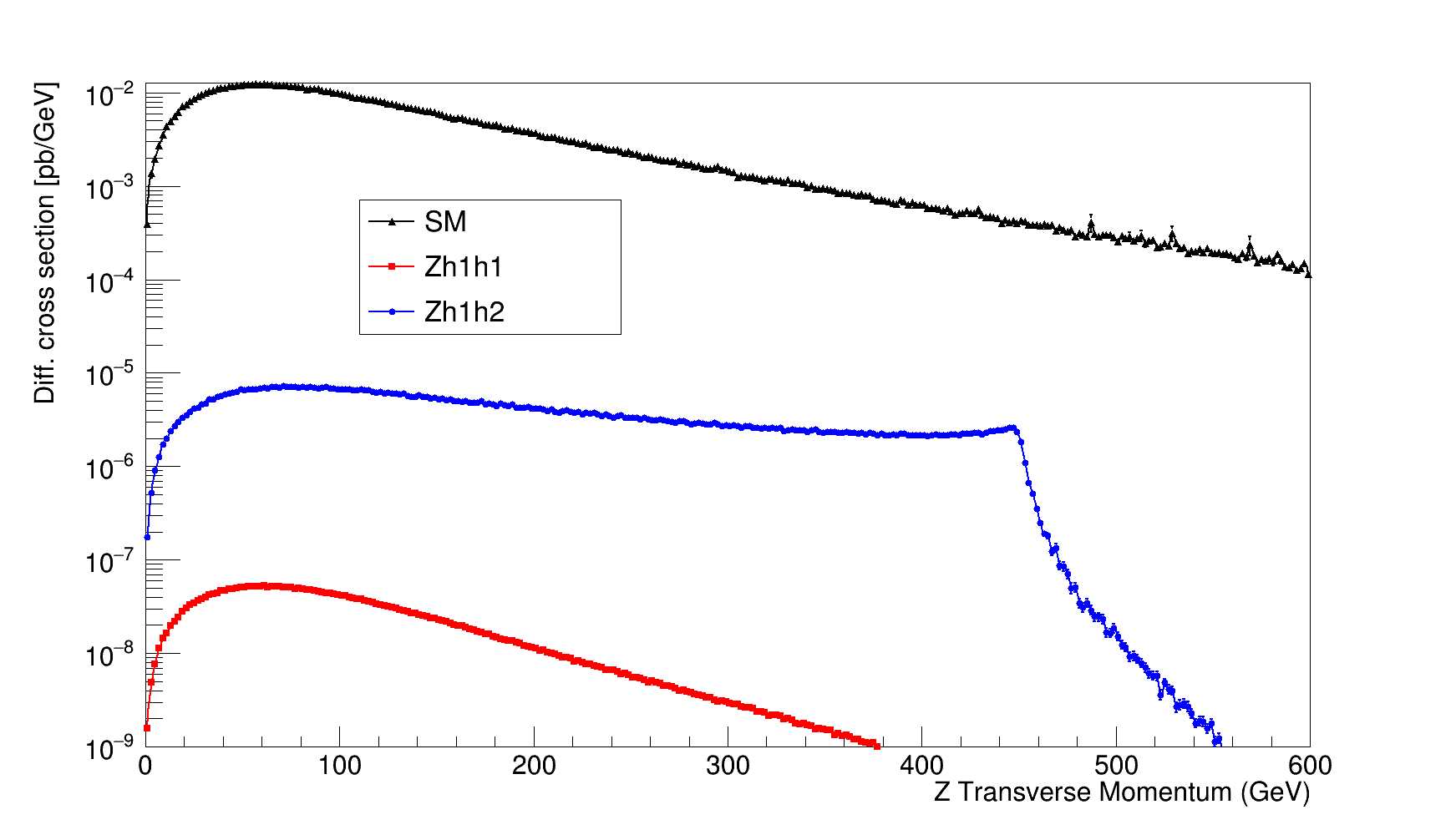}
\begin{center}
a)
\par\end{center}%
\end{minipage}%
\begin{minipage}[t]{0.5\columnwidth}%
\includegraphics[scale=0.11]{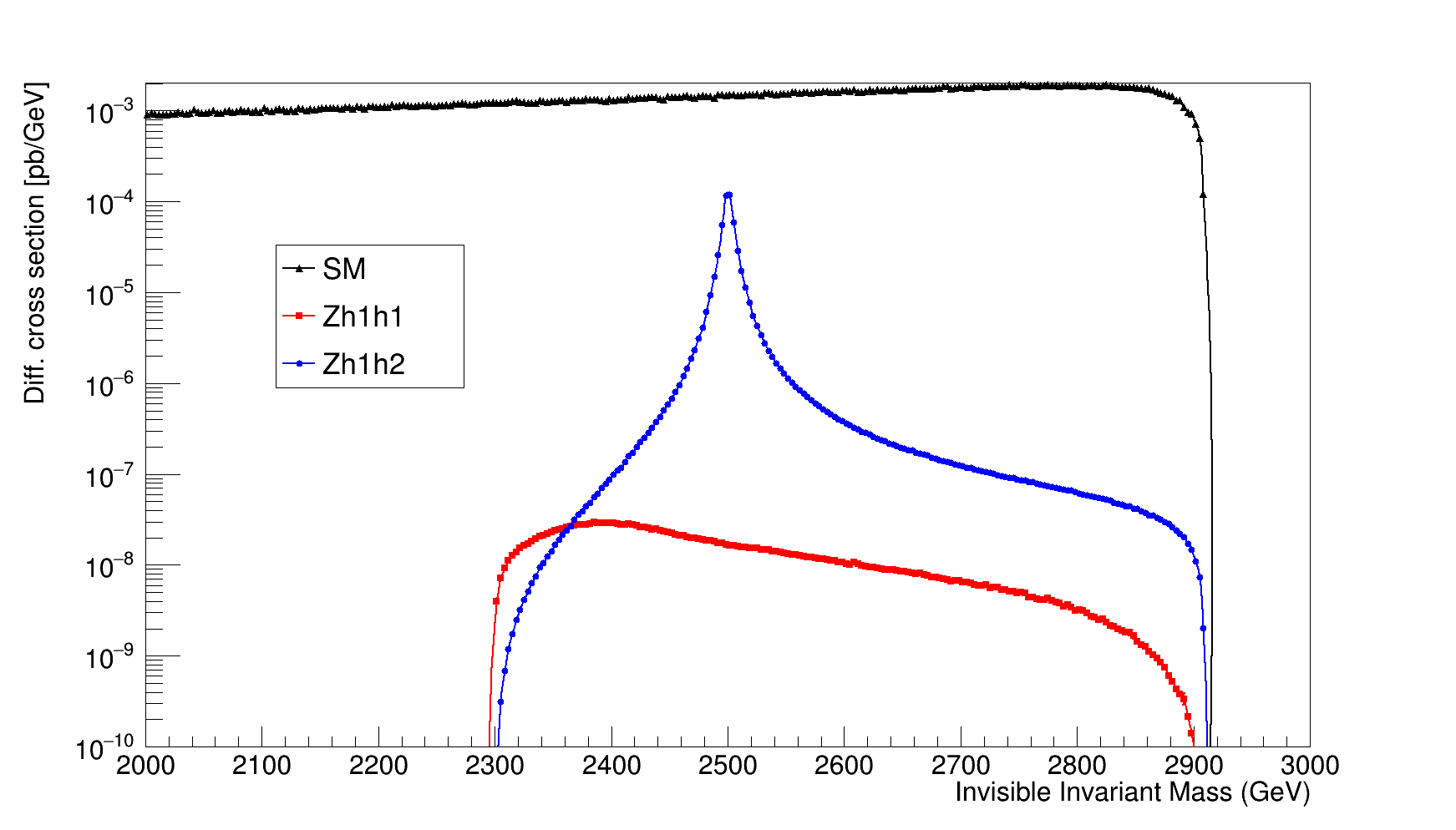}
\begin{center}
b)
\par\end{center}%
\end{minipage}

\caption{$P_{T}$ and missing invariant mass distributions for $e{{}^+}e{{}^-}\rightarrow Zh_{1}h_{1}$
and $e\text{\textsuperscript{+}}e\text{\textsuperscript{-}}\rightarrow Zh_{1}h_{2}$
with $M_{\rho}=2.5$ TeV. For comparison purposes we include the distributions
for $e{{}^+}e{{}^-}\rightarrow Z\nu_{e}\bar{\nu}_{e}$predicted by
the Standard Model (labeled ``SM'' in the pictures). }

\label{Fig:MonoZDist}

\end{figure}
In Figure \ref{Fig:MonoZDist} we show the transverse momentum distribution
of the $Z$ boson (\ref{Fig:MonoZDist} a) and the missing invariant
mass distribution (\ref{Fig:MonoZDist} b) for $e{{}^+}e{{}^-}\rightarrow Zh_{1}h_{1}$
and $e\text{\textsuperscript{+}}e\text{\textsuperscript{-}}\rightarrow Zh_{1}h_{2}$
with $M_{\rho}=2.5$ TeV without any cut. For comparison purposes
we include the SM distributions for $e{{}^+}e{{}^-}\rightarrow Z\nu_{e}\bar{\nu}_{e}$
which is the main source of background. We see that although the main
component of the signal ($Zh_{1}h_{2}$ ) has a non-trivial structure
due to the resonant contribution of the $\rho_{0}$ in the s-channel,
the signal is in general several orders of magnitude smaller than the
background. When we apply the cuts $P_{TZ}\geqq100$ GeV and $M_{\rho}-100\mathrm{GeV}\leqq M_{\mathrm{missing}}\leqq M_{\rho}+100\mathrm{GeV}$
(where $P_{TZ}$ is the transverse momentum of the $Z$ boson and
$M_{\mathrm{missing}}$ is the missing invariant mass) we obtain:

\[
\sigma^{\mathrm{signal}}=\sum_{i=1}^{2}\sigma\left(e{^{+}}e{^{-}}\rightarrow Zh_{1}h_{i}\right)\approx\left(1.6\pm0.2\right)\mathrm{fb}
\]
almost independently of the value of $M_{\rho}$. On the other hand,
the background cross section, under the same kinematic cuts, results
to be:

\[
\sigma^{\mathrm{back}}=\left(1.4\pm0.3\right)\times10^{2}\mathrm{fb}.
\]

After the cuts, the signal is still two orders of magnitude smaller than
the background. However with a luminosity of $\mathcal{L}=2000\,\mathrm{fb}^{-1}$
the event excess will represent a statistical significance of $S=6$. This significance, however,
reflects only the statistical uncertainty. A more detailed analysis should take into account the 
systematic uncertainties coming form the particular characteristics of the detector and the 
reconstruction of the $Z$ boson specially from its hadronic decay. An additional difficulty is represented by the invisible decay of the $Z$ boson into neutrinos.

\section{Conclusions\label{sec:Conclusions}}

In this paper, we studied the possibility of using CLIC to discover
a new vector resonance associated to a hypothetical strongly coupled
dark sector. We focused on the di-muon production and we have strongly
relied on the radiative return to the resonance. In the case where
the non-standard scalars are heavy enough to forbid the decay of the
vector resonances into particles of the dark sector, we predict an
important excess of event for the resonance masses we considered,even
when a moderate smearing of the momentum is taken into account. When
the scalars are light, the new vector become broad and it is difficult
to define a resonance. A remarkable exception is found when the mass
of the scalars is close to (but smaller than) a half of the resonance
mass and the scale of the dark sector is high. In all these positive
cases, the high energy and high luminosity projected for CLIC are
enough for leading to the discovery of the new dark resonance or severely
constrain the model. However, for the light scalar case, the pair production 
of charged scalars arises as an attractive alternative for discovering the dark sector. 
In a complementary analysis, we studied the mono-$Z$
production. We found that, although this channel is more challenging
that the di-muon production, after the implementation of an adequate
cut on the missing invariant mass, the high luminosity regime of CLIC
can lead to an statistically significant excess of events. These facts
suggest that CLIC would be an excellent environment for testing models
with a complex dark sector and illustrate the importance of building
such a machine for testing models beyond the Standard Model.

\section*{Acknowledgments}

This work was supported in part by Conicyt (Chile) grants PIA/ACT-1406
and PIA/Basal FB0821, and by Fondecyt (Chile) grant 1160423. AZ is
very thankful to the developers of MAXIMA \cite{Maxima} and the
package Dirac2 \cite{Dirac2} .These packages were used in parts
of this work.


\begin{thebibliography}{99}
\bibitem{i2HDM-1}Nilendra G. Deshpande and Ernest Ma, \textquotedbl Pattern
of Symmetry Breaking with Two Higgs Doublets\textquotedbl , Phys.Rev.
D18 (1978), pp. 2574.

\bibitem{i2HDM-2}L. Lopez Honorez, E. Nezri, J. F. Oliver and M.
H. G. Tytgat, ``The Inert Doublet Model: An Archetype for Dark Matter,''
JCAP \textbf{0702}, 028 (2007) doi:10.1088/1475-7516/2007/02/028 {[}hep-ph/0612275{]}.

\bibitem{i2HDM-3}R. Barbieri, L. J. Hall and V. S. Rychkov, ``Improved
naturalness with a heavy Higgs: An Alternative road to LHC physics,''
Phys. Rev. D \textbf{74}, 015007 (2006) doi:10.1103/PhysRevD.74.015007
{[}hep-ph/0603188{]}.

\bibitem{VRi2HDM}F. Rojas-Abatte, M. L. Mora, J. Urbina and A. R.
Zerwekh, ``Inert two-Higgs-doublet model strongly coupled to a non-Abelian
vector resonance'' Phys. Rev. D \textbf{96}, no. 9, 095025 (2017)
doi:10.1103/PhysRevD.96.095025 {[}arXiv:1707.04543 {[}hep-ph{]}{]}.

\bibitem{Strom:2017pjx=00007D}R. Ström {[}CLICdp Collaboration{]},
``Overview of the CLIC detector and its physics potential,'' EPJ Web
Conf. \textbf{164}, 01020 (2017). doi:10.1051/epjconf/201716401020

\bibitem{Abramowicz:2016zbo}H. Abramowicz \textit{et al}., ``Higgs
physics at the CLIC electron--positron linear collider,'' Eur. Phys.
J. C \textbf{77}, no. 7, 475 (2017) doi:10.1140/epjc/s10052-017-4968-5
{[}arXiv:1608.07538 {[}hep-ex{]}{]}.

\bibitem{Abramowicz:2018rjq}H.\textasciitilde Abramowicz \textit{et
al}.,{[}CLICdp Collaboration{]}, ``Top-Quark Physics at the CLIC Electron-Positron
Linear Collider,'' arXiv:1807.02441 {[}hep-ex{]}.

\bibitem{Milutinovic-Dumbelovic:2018drv=00007D}G. Milutinovic-Dumbelovic
{[}CLICdp Collaboration{]}, ``Higgs and BSM physics at CLIC,'' PoS
EPS \textbf{-HEP2017}, 319 (2018). doi:10.22323/1.314.0319

\bibitem{vanderKolk:2017urn}N. van der Kolk {[}CLICdp Collaboration{]},
``CLIC Detector and Physics Status,'' arXiv:1703.08876 {[}physics.ins-det{]}.

\bibitem{Simoniello:2016veu}R. Simoniello {[}CLICdp Collaboration{]},
``BSM physics at CLIC,'' PoS ICHEP \textbf{2016}, 153 (2016). doi:10.22323/1.282.0153

\bibitem{Limit-LHC-1}O. Castillo-Felisola, C. Corral, M. González,
G. Moreno, N. A. Neill, F. Rojas, J. Zamora and A. R. Zerwekh, ``Higgs
Boson Phenomenology in a Simple Model with Vector Resonances,'' Eur.\textbackslash{}
Phys.\textbackslash{} J.\textbackslash{} C \{\textbackslash bf 73\},
no. 12, 2669 (2013) doi:10.1140/epjc/s10052-013-2669-2 {[}arXiv:1308.1825
{[}hep-ph{]}{]}.

\bibitem{Limit-LHC-2}A. E. Carcamo Hernandez, C. O. Dib and A. R.
Zerwekh, ``The Effect of Composite Resonances on Higgs decay into
two photons,'' Eur.\textbackslash{} Phys.\textbackslash{} J.\textbackslash{}
C \{\textbackslash bf 74\}, 2822 (2014) doi:10.1140/epjc/s10052-014-2822-6
{[}arXiv:1304.0286 {[}hep-ph{]}{]}.

\bibitem{Limit-LHC-3}M. Gintner and J. Juran, ``The LHC mass limits
for the \$SU(2)\_\{L+R\}\$ vector resonance triplet of a strong extension
of the Standard model,'' Acta Phys.\textbackslash{} Polon.\textbackslash{}
B \{\textbackslash bf 48\}, 1383 (2017) doi:10.5506/APhysPolB.48.1383
{[}arXiv:1705.04806 {[}hep-ph{]}{]}.

\bibitem{CalcHEP}A. Belyaev, N. D. Christensen and A. Pukhov, ``CalcHEP
3.4 for collider physics within and beyond the Standard Model,''
Comput. Phys. Commun. \textbf{184}, 1729 (2013) doi:10.1016/j.cpc.2013.01.014
{[}arXiv:1207.6082 {[}hep-ph{]}{]}.

\bibitem{LanHEP-1}A. Semenov, ``LanHEP: A Package for the automatic
generation of Feynman rules in field theory. Version 3.0,'' Comput.
Phys. Commun. \textbf{180}, 431 (2009) doi:10.1016/j.cpc.2008.10.012
{[}arXiv:0805.0555 {[}hep-ph{]}{]}.

\bibitem{LanHEP-2}A. Semenov, ``LanHEP --- A package for automatic
generation of Feynman rules from the Lagrangian. Version 3.2,'' Comput.
Phys. Commun. \textbf{201}, 167 (2016) doi:10.1016/j.cpc.2016.01.003
{[}arXiv:1412.5016 {[}physics.comp-ph{]}{]}.

\bibitem{PDG}C. Patrignani et al. (Particle Data Group), Chin. Phys.
C, 40, 100001 (2016) and 2017 update. 

\bibitem{Maxima}Maxima, \textquotedblleft Maxima, a computer algebra
system. version 5.40.0\textquotedblright , 2017. http://maxima.sourceforge.net/.

\bibitem{Dirac2}E. L. Woollett, \textquotedblleft Dirac2: A high
energy physics package for maxima\textquotedblright , 2012. http://web.csulb.edu/\textasciitilde woollett/.
\end{thebibliography}
\end{document}